\documentclass{article}
\usepackage[utf8]{inputenc}
\usepackage[T1]{fontenc}
\usepackage{algorithm, algorithmic}
\usepackage{microtype}
\usepackage{graphicx}
\usepackage{subfigure}
\usepackage{booktabs} 
\usepackage{hyperref}
\usepackage{times}
\usepackage{url}
\usepackage[english]{babel}
\usepackage{ulem}
\usepackage{amsmath}
\usepackage{amsthm}
\usepackage{amsfonts}
\usepackage{amssymb}
\usepackage{bm}
\usepackage{float}
\usepackage{graphicx}
\usepackage{nicefrac}
\usepackage{soul}
\usepackage{stmaryrd}
\usepackage{comment}
\usepackage{mathtools}
\usepackage[makeroom]{cancel}
\usepackage{tcolorbox}
\usepackage[capitalize,noabbrev]{cleveref}

\usepackage[parfill]{parskip}

\author{
    Antoine Collas \\
    Université Paris-Saclay, Inria, CEA
    \and Rémi Flamary \\
    Ecole Polytechnique, Institut Polytechnique de Paris, CMAP, UMR 7641
    \and Alexandre Gramfort \\
    Université Paris-Saclay, Inria, CEA
}
\date{}

\theoremstyle{plain}
\newtheorem{theorem}{Theorem}[section]
\newtheorem{proposition}[theorem]{Proposition}
\newtheorem{corollary}[theorem]{Corollary}

\usepackage[textsize=tiny]{todonotes}

\title{Weakly supervised covariance matrices alignment through Stiefel matrices estimation for MEG applications}

\newcommand{\overbar}[1]{\mkern 1.5mu\overline{\mkern-1.5mu#1\mkern-1.5mu}\mkern 1.5mu}

\newcommand{\norm}[1]{\left\lVert#1\right\rVert}

\DeclareMathOperator*{\argmin}{arg\,min \,}

\DeclareMathOperator{\diag}{diag}

\let\hbar\relax
\newcommand{\hbar}{\overbar{h}}

\DeclareMathOperator*{\minimize}{minimize \,}

\DeclareMathOperator{\Prob}{\mathbb{P}}

\DeclareMathOperator{\sign}{sign}
\DeclareMathOperator{\spann}{span}

\DeclareMathOperator*{\Tr}{Tr}

\DeclareMathOperator{\vect}{vec}

\newcommand{\domain}{{\mathcal{D}}}
\newcommand{\source}{{\mathcal{S}}}
\newcommand{\target}{{\mathcal{T}}}

\newcommand*{\VEC}[1]  {\bm{#1}}
\newcommand*{\MAT}[1]  {\bm{#1}}

\newcommand{\bA}{{\MAT{A}}}

\newcommand{\bbeta}{\VEC{\beta}}

\newcommand{\bC}{\MAT{C}}

\newcommand{\bE}{\MAT{E}}
\newcommand{\bEMean}{\overbar{\bE}}

\newcommand{\bEta}{{\VEC{\eta}}}

\newcommand{\bGamma}{\MAT{\Gamma}}

\newcommand{\bI}{{\MAT{I}}}

\newcommand{\bK}{\MAT{K}}

\newcommand{\bn}{\VEC{n}}

\newcommand{\bN}{\MAT{N}}
\newcommand{\bO}{\MAT{O}}

\newcommand{\bp}{\VEC{p}}

\newcommand{\bPi}{\MAT{\Pi}}
\newcommand{\bpi}{\MAT{\pi}}
\newcommand{\bQ}{\MAT{Q}}

\newcommand{\bSigma}{{\MAT{\Sigma}}}
\newcommand{\bSigmaMean}{{\overbar{\bSigma}}}

\newcommand{\bU}{{\MAT{U}}}

\newcommand{\bW}{\MAT{W}}

\newcommand{\bx}{{\VEC{x}}}

\newcommand{\bX}{\MAT{X}}

\newcommand{\bXi}{{\VEC{\xi}}}
\newcommand{\by}{{\VEC{y}}}

\newcommand{\bzero}{\VEC{0}}
\newcommand{\bone}{\VEC{1}}

\newcommand{\Ort}{{\mathcal{O}}}

\newcommand{\R}{\mathbb{R}}

\newcommand{\Spos}{{\mathbb{S}_p^{++}}}

\newcommand{\St}{{\textup{St}(d,q)}}
\newcommand{\Sym}{{\mathbb{S}_p}}

\newcommand{\bpbar}{\bar{\VEC{p}}}
\newcommand{\pbar}{\bar{p}}

\newcommand{\loss}{\mathcal{L}}
\newcommand{\lossSt}{\mathcal{L}^*}

\newcommand{\losssup}{\loss_\text{sup.}}
\newcommand{\lossemb}{\loss_\text{met.}}

\newcommand{\lossgrass}{\loss_\text{Gr}}
\newcommand{\cOT}{\gamma}
\newcommand{\cgrass}{\rho}
\newcommand{\csup}{\varepsilon}

\begin{document}
\maketitle

\begin{abstract}
    This paper introduces a novel domain adaptation technique for time series data, called Mixing model Stiefel Adaptation (MSA), specifically addressing the challenge of limited labeled signals in the target dataset.
    Leveraging a domain-dependent mixing model and the optimal transport domain adaptation assumption, we exploit abundant unlabeled data in the target domain to ensure effective prediction by establishing pairwise correspondence with equivalent signal variances between domains.
    Theoretical foundations are laid for identifying crucial Stiefel matrices, essential for recovering underlying signal variances from a Riemannian representation of observed signal covariances.
    We propose an integrated cost function that simultaneously learns these matrices, pairwise domain relationships, and a predictor, classifier, or regressor, depending on the task.
    Applied to neuroscience problems, MSA outperforms recent methods in brain-age regression with task variations using magnetoencephalography (MEG) signals from the Cam-CAN dataset.
\end{abstract}

\section{Introduction}
    \label{sec:intro}
    \begin{figure*}[t]
    \centering
    \includegraphics[width=\linewidth]{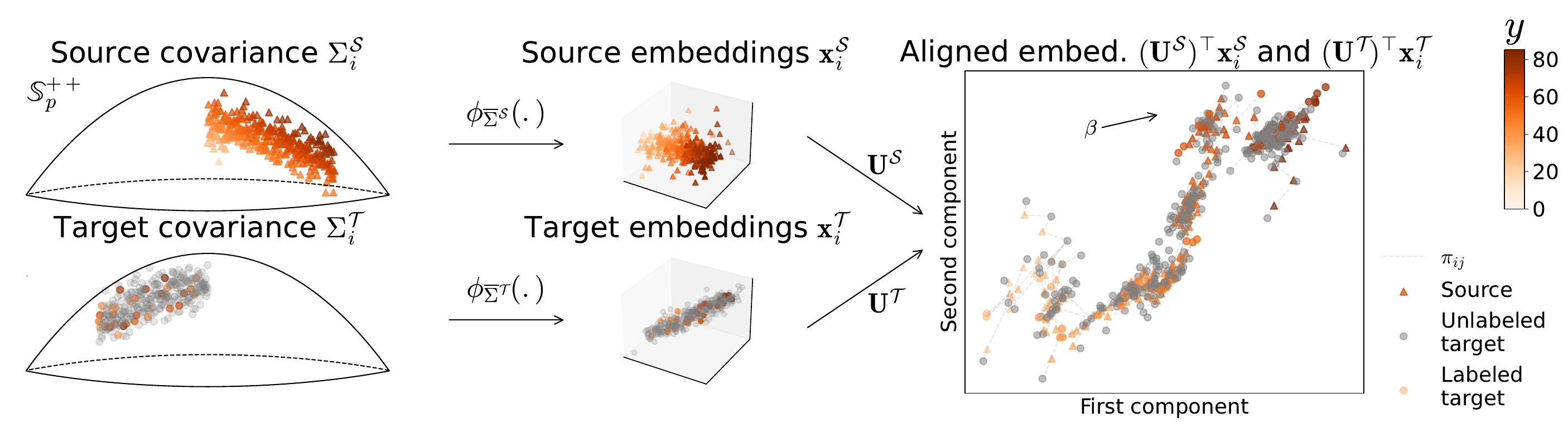}
    \caption{
        \textbf{Illustration of MSA for a regression task.}
        Source and target covariance matrices, $\bSigma_i^\source$ and $\bSigma_i^\target$, exhibit different patterns in their original spaces.
        Embedding them into $\bx_i^\source$ and $\bx_i^\target$ and then jointly learning optimal transport plan $\bpi$ and orthogonal bases $\bU^\source$ and $\bU^\target$ alleviate this problem by finding components that matter for prediction.
    }
\end{figure*}

Multivariate time series data is ubiquitous in a wide range of applications, such as remote sensing, finance, and neuroscience.
In numerous scenarios, observed time series result from linear combinations of underlying signals of interest.
In remote sensing, measured spectral bands are linear mixtures of endmember spectra (e.g., water, grass, wood) with coefficients corresponding to the proportions of these endmembers within the observed scene~\cite{keshava_spectral_2002}.
In neuroscience, a suitable generative model for magnetoencephalography (MEG) and electroencephalography (EEG) data is the linear instantaneous mixing model, leveraging the inherent linearity of Maxwell's equations~\cite{hamalainen_magnetoencephalographytheory_1993}.

Covariance matrices have emerged as powerful descriptors for diverse data types, offering valuable insights into the underlying relationships within multivariate data.
The application of Riemannian geometry to covariance matrices~\cite{skovgaard_riemannian_1984} has paved the way for innovative solutions in tasks such as classification~\cite{tuzel_pedestrian_2008, huang_riemannian_2017} and regression~\cite{sabbagh_manifold-regression_2019, sabbagh_predictive_2020, bonet_sliced-wasserstein_2023}.
These models have been successfully employed in the analysis of multivariate time series data, including applications in remote sensing~\cite{collas_riemannian_2023} and M/EEG~\cite{barachant_multiclass_2012, yger_riemannian_2017, kobler_spd_2022} applications.
Notably, recent studies have revealed the competitiveness of shallow covariance-based models when compared to deep learning techniques in neuroscience~\cite{engemann_reusable_2022}.

However, the challenge of domain adaptation poses a substantial hurdle in the analysis of multivariate time series data.
For example, when data are measured using different sensors or at different periods, domain shifts occur, making it difficult to apply classifiers or regressors across domains directly.
Domain adaptation addresses these shifts by aligning data~\cite{courty_optimal_2017, kouw_review_2021}. 
This challenge is particularly pronounced in M/EEG, where changes in recording devices, head morphology, experimental protocol, or data collection sites can induce shifts.
It is especially crucial in applications like brain-computer interfaces (BCI)~\cite{jayaram_transfer_2016, zanini_transfer_2018, maman_domain_2019, rodrigues_riemannian_2019, bleuze_transfer_2021} and brain age prediction~\cite{mellot_harmonizing_2023}, where limited labeled data in the target domain necessitates effective adaptation techniques.

In this paper, we propose Mixing model Stiefel Adaptation (MSA), a novel domain adaptation approach for time series data under mixing models.
We address the challenging problem of adapting models when the target domain has limited labeled data while a fully labeled source domain is available.
Our approach involves the identification of two Stiefel matrices that are
used to align the covariances of both source and target data.
When applied to a Riemannian representation of the covariances of observed signals, these matrices facilitate the recovery of underlying signal variances.
To achieve this, MSA simultaneously
learns these matrices and a predictor.
The proposed framework is general enough to be applied to both classification and regression tasks.

In the subsequent sections, we provide a detailed exposition of our novel approach.
We show that MSA outperforms recent approaches in brain-age regression using
magnetoencephalography (MEG) signals from the Cam-CAN dataset~\cite{taylor_cambridge_2017} while accommodating task variations.

\section{Domain shift and covariance matrices under mixing models}
	\label{sec:problem}
	
This section presents time series under mixing models in the context of domain adaptation with mixing matrices that differ from one domain to another.
These data are summarized with covariance matrices as discussed in the
introduction.
Then, the ``optimal transport domain adaptation (OTDA) assumption" is introduced: we assume that, for large enough domains, there exists a permutation matrix that associates source variances with target variances. 
Furthermore, the main tools of Riemannian geometry associated with covariance matrices are presented.

\subsection{Mixing models in a domain adaption context}
\label{subsec:mixing}
Domain adaptation is a core machine learning technique used to enhance model performance when there is a distribution shift between the source and target domains denoted $\source$ and $\target$, respectively.
When the source domain provides labeled data for training, and only a fraction of
the target domain is labeled, the problem is known as semi-supervised domain adaptation.

\textbf{Domain specific mixing model}
We consider the covariance matrices of the signal as features for regression and classification.
We assume that the signal is the sum of a mixed signal of interest plus an independent noise.
Hence, for any domain $\domain \in \{\source, \target\}$, the covariance matrix of the $i^\text{th}$ signal is
\begin{equation}
    \label{eq:mixing_model_compact}
    \bSigma_i^\domain \triangleq \bA^\domain
    \begin{bmatrix}
        \diag(\bp_i^\domain) & \bzero_{q \times (p-q)} \\
        \bzero_{(p-q) \times q} & \bN_i^\domain
    \end{bmatrix}
    (\bA^\domain)^\top \in \R^{p \times p}
\end{equation}
where $\bA^\domain$ is the domain-specific mixing matrix, $\bp_i^\domain = [p_{i,1}^\domain, \dots, p_{i,q}^\domain]^\top \in \R^{q}$ contains the variances of the unmixed signal of interest, $\bzero$ is the zero matrix and $\bN_i^\domain \in \R^{(p-q) \times (p-q)}$ is the covariance matrix of the noise.

\subsection{Covariance matrices in the Riemannian geometry framework}
\label{subsec:riemann}

We utilize the Riemannian geometry of symmetric positive definite matrices for both regressing outcomes and classifying covariance matrices.
This approach is theoretically grounded, as supported by studies such as~\cite{skovgaard_riemannian_1984, pennec_riemannian_2006}.
In practical applications, the use of Riemannian geometry for covariance matrices demonstrates strong performance, notably in EEG~\cite{yger_riemannian_2017, bonet_sliced-wasserstein_2023} and MEG~\cite{engemann_reusable_2022, mellot_harmonizing_2023} applications.

\textbf{Riemmanian geometry of $\Spos$}
\sloppy
The covariance matrices belong to the set of $p\times p$ symmetric positive definite matrices denoted $\Spos$.
The latter is open in the set of $p\times p$ symmetric matrices denoted $\Sym$, and thus is a smooth manifold~\cite{boumal_introduction_2023}.
A vector space is defined at each $\bSigma \in \Spos$, called the tangent space and denoted $T_\bSigma \Spos$.
It is equal to $\Sym$, the ambient space.
Equipped with a smooth inner product, a smooth manifold becomes a Riemannian manifold.
To do so, we leverage the affine invariant Riemannian metric~\cite{skovgaard_riemannian_1984}.
Indeed, it gives good theoretical properties to the $\Spos$ manifold such as being geodesically complete and works well in practice~\cite{barachant_multiclass_2012, yger_riemannian_2017, bonet_sliced-wasserstein_2023}.
Given $\bGamma, \bGamma^\prime \in \Sym$, this metric writes $\langle \bGamma, \bGamma^\prime \rangle_\bSigma = \Tr\left(\bSigma^{-1} \bGamma \bSigma^{-1} \bGamma^\prime \right)$ and the corresponding norm is $\norm{\bGamma}_\bSigma = \sqrt{\langle \bGamma, \bGamma \rangle_\bSigma}$.

\textbf{Riemannian mean on $\Spos$}
The Riemannian distance (or geodesic distance) associated with the affine invariant metric is
$d_\Spos(\bSigma, \bSigma^\prime) = \norm{\log\left(\bSigma^{\nicefrac{-1}{2}} \bSigma^\prime \bSigma^{\nicefrac{-1}{2}}\right)}_F$
with $\log: \Spos \to \Sym$ being the matrix logarithm.
It should be noted that, since the Riemannian metric is affine invariant, $d_{\Spos}$ is also affine invariant, i.e. for every $\bA \in \R^{p \times p}$ invertible, we have $d_{\Spos}(\bA \bSigma \bA^{\top}, \bA \bSigma^\prime \bA^{\top}) = d_{\Spos}(\bSigma, \bSigma^\prime)$.
In the following, we vectorize covariance matrices $\bSigma_i^\domain$ in a tangent space $T_\bSigmaMean\Spos$.
A classical choice for the base point is the Riemannian mean $\bSigmaMean$. This mean, denoted as $\bSigmaMean$, is defined for a set $\bSigma_1, \dots, \bSigma_n \in \Spos$ as
\begin{equation}
    \label{eq:mean}
    \bSigmaMean \triangleq \argmin_{\bSigma \in \Spos} \sum_{i=1}^n d_{\Spos}(\bSigma, \bSigma_i)^2
\end{equation}
and is computed with a Riemannian gradient descent~\cite{pennec_riemannian_2006, zhang_first-order_2016}.

\textbf{Linearization of the $\Spos$ manifold}
The affine invariant metric induces the Riemannian exponential $\exp_\bSigma: \Sym \to \Spos$
which is such that, for $t\in[0,1]$, $\exp_\bSigma(t\bGamma)$ is the geodesic
with initial position $\bSigma$ and speed $\bGamma$.
The inverse operator is the Riemannian logarithm $\log_\bSigma: \Spos \to \Sym$.
This operator is preponderant in the rest of the paper since it transforms covariance matrices that belong to a Riemannian manifold into elements of $\Sym$, a vector space.
Furthermore, the Riemannian logarithm can be vectorized using the operator $\phi_\bSigma$ defined such that $\norm{\log_\bSigma(\bSigma^\prime)}_\bSigma = \norm{\phi_\bSigma(\bSigma^\prime)}_2$ and that is
\begin{equation}
    \label{eq:vectorize}
    \phi_\bSigma(\bSigma^\prime) \triangleq \vect\left( \log\left(\bSigma^{\nicefrac{-1}{2}} \bSigma_i^\prime \bSigma^{\nicefrac{-1}{2}}\right) \right).
\end{equation}
Depending on the context, the operator $\vect$ is either the full vectorization (hence $\phi_\bSigma(\bSigma^\prime)  \in \R^{p^2}$) or the vectorization of the upper triangular part with off-diagonal elements multiplied by $\sqrt{2}$ to preserve the norm (hence $\phi_\bSigma(\bSigma^\prime)  \in \R^{p(p+1)/2}$).

\textbf{Riemannian geometry for domain adaptation}
For any $\domain \in \{\source, \target\}$, we denote the source and target embeddings by
\begin{equation}
    \label{eq:embedding}
    \bx_i^\domain \triangleq \phi_{\bSigmaMean^\domain}(\bSigma_i^\domain)
\end{equation}
where $\bSigmaMean^\domain$ is the Riemannian mean of the $\bSigma_i^\domain$.
Finally, it should be noted that, due to the equation~\eqref{eq:pi} and the affine invariance of the Riemannian distance, we have that
\begin{equation}
    \bSigmaMean^\domain = \bA^\domain \bEMean^\domain (\bA^\domain)^{\top}
\end{equation}
where $\bEMean^\domain$ is a block diagonal matrix with the upper left block equal to $\diag(\bpbar) \in \R^{q \times q}$ with elements $\pbar_l = (\prod_{i=1}^n p_{i, l}^\source)^{\nicefrac{1}{n}} = (\prod_{i=1}^n p_{il}^\target)^{\nicefrac{1}{n}}$.
 
\subsection{Regression and classification with mixed and domain shifted signals}
\label{subsec:models_reg_classif}

We consider two classical models in M/EEG on the decision function for regression and classification problems.
These models relate outcomes $y_i^\domain$ with the variances $\{p_{il}^\domain\}_{l=1}^q$~\cite{grosse-wentrup_multiclass_2008, blankertz_optimizing_2008, dahne_spoc_2014}.
We extend these models to domain adaptation, and notably, we assume parameters $\beta_0, \dots, \beta_q \in \R$ of these decision functions to be equal between source and target domains.

\textbf{Predicition models: regression and classification}
For regression problems, a classical assumption is the existence of a log-linear relationship between the value to regress $y_i^\domain$ of the $i^\text{th}$ subject in domain $\domain$ and the associated variances $p_{il}^\domain$~\eqref{eq:mixing_model_compact}, i.e., there exists $\beta_0, \dots, \beta_q$ such that
\begin{equation}
    \label{eq:model_reg}
    y_i^\domain = \sum_{l=1}^{q} \beta_l \log(p_{il}^\domain) + \beta_0 + \varepsilon_i^\domain
\end{equation}
where $\varepsilon_i^\domain$ is a random noise.
Considering a binary classification problem\footnote{Throughout the paper, classification problems are presented with two classes only to simplify the exposition. The extension to multiple classes is straightforward.} with outcomes in $\{-1, 1\}$, the second model assumes that $y_i^\domain$ is the sign of a log-linear regression on $\{p_{il}^\domain\}_{l=1}^q$, i.e.,
\begin{equation}
    \label{eq:model_classif}
    y_i^\domain = \sign\left( \sum_{l=1}^{q} \beta_l \log(p_{il}^\domain) + \beta_0 + \varepsilon_i^\domain \right)
\end{equation}
where $\sign(a) = 1$ if $a \geq 0$ and $-1$ otherwise.

\textbf{OTDA assumption}
An additional assumption is the optimal transport domain adaptation (OTDA) assumption.
In our setup, we assume there exists a matrix $\bpi \in \bPi(n, m) \triangleq \left\{ \bpi \in \R^{n \times m}: \pi_{ij} \geq 0, \bpi \bone_m = \frac{1}{n} \bone_n, \bpi^{\top} \bone_n = \frac{1}{m} \bone_m \right\}$, the set of discrete joint distributions of uniform marginals, such that
\begin{equation}
    \label{eq:pi}
    \bp_i^\source = \bp_j^\target \quad \text{if} \quad \pi_{ij} > 0.
\end{equation}
This assumption is tightly related to the assumption of preserved conditional distribution stated in~\cite{courty_optimal_2017}.
Indeed, the authors assume the label information is preserved by the transformation $\bpi$, i.e., for every $i$, $\Prob_\source(y | \bp_i^\source) = \Prob_\target(y | m(\bp_i^\source))$ with $\Prob_\domain$ the conditional probability distribution on the domain $\domain$ and $m(\bp_i^\source)^\top = (\bpi [\bp_1^\target, \cdots, \bp_m^\target]^\top)_{[i, :]}$.
This assumption holds in our setup if the noise $\varepsilon_i^\domain$ has the same distribution for the two domains.

\textbf{Prediction from covariance matrices}
These models motivate the recovery of $\bp_i^\domain$ from $\bx_i^\domain$.
The following proposition indicates that, under the mixing model~\eqref{eq:mixing_model_compact}, this recovery is possible through orthogonal projections of $\bx_i^\source$ and $\bx_j^\target$ with well-chosen matrices.
The latter belongs to the Stiefel manifold, the set of $q$-dimensional orthogonal basis in $\R^{p^2}$ denoted
$\text{St}(p^2, q) \triangleq \{ \bU\in \R^{p^2 \times q}: \bU^{\top}\bU= \bI_q \}$.

\begin{proposition}[Stiefel projections]
    \label{prop:projections}
    Given source and target embeddings, $\bx_i^\source$ and $\bx_j^\target$~\eqref{eq:embedding}, following mixing models~\eqref{eq:mixing_model_compact}, with $\pi_{ij} > 0$~\eqref{eq:pi}, there exist $\bU^\source, \bU^\target \in \textup{St}(p^2, q)$ such that
    \begin{equation*}
        (\bU^\source)^{\top} \bx_i^\source   \!=\! (\bU^\target)^{\top} \bx_j^\target   \!=\!   \left[\log(p_{i, 1}/\pbar_1), \dots, \log(p_{i, q}/\pbar_q) \right]^{\top}
    \end{equation*}
    where $p_{i,l} \triangleq p_{i,l}^\source = p_{j,l}^\target$, $\forall l \in \llbracket 1, q \rrbracket$,
\end{proposition}
The proof is available in the Appendix~\ref{app:proof}.
Intuitively, $\bU^\source$ and $\bU^\target$ extract the signal of interest and unmix it, i.e., invert $\bA^\source$ and $\bA^\target$.
Thus, these Stiefel matrices remove the shift between source and target domains.
Combining Proposition~\ref{prop:projections} with models~\eqref{eq:model_reg} and~\eqref{eq:model_classif} gives two new models that relate $\bx_i^\domain$ with $y_i^\domain$.
\begin{corollary}[Realined predictive models]
    \label{corollary:models}
    For all $\domain \in \{\source, \target\}$ and under mixing models~\eqref{eq:mixing_model_compact}, there exists $\bU^\domain \in \textup{St}(p^2,q)$ and $\beta_0, \cdots, \beta_q \in \R$ such that the data $(\bx_i^\domain, y_i^\domain)$ follow the models:
    \begin{equation*}
        y_i^\domain  = \ \sum_{l=1}^q \beta_l \left((\bU^\domain)^{\top} \bx_i^\domain\right)_l + \beta_0 + \varepsilon_i^\domain,
    \end{equation*}
    in regression, and
    \begin{equation*}
        y_i^\domain  =  \sign  \left( \sum_{l=1}^q \beta_l \left((\bU^\domain)^{\top} \bx_i^\domain\right)_l + \beta_0 + \varepsilon_i^\domain  \right)
    \end{equation*}
    in classification.
\end{corollary}

We finish this section by pointing out that until now, $\bx_i^\domain \in \R^{p^2}$.
In practice, we use the half vectorization as explained in section~\ref{subsec:riemann} to reduce the dimension of the embeddings to $d \triangleq p(p+1)/2$.

\section{Joint learning of \texorpdfstring{$\bU^\source$}{TEXT}, \texorpdfstring{$\bU^\target$}{TEXT} and \texorpdfstring{$\bbeta$}{TEXT}}
    \label{sec:cost_optim}
	
In the last section, we defined our data, their underlying mixing model, and the domain adaptation setup.
We now present an algorithm to learn $\bU^\source$, $\bU^\target$ and $\bbeta = [\beta_0, \cdots, \beta_q] \in \R^{q+1}$ from data.

\textbf{Setup}
We assume having access from a fully labeled source domain and a partially labeled target domain.
Indeed, $n$ source data, $\{(\bx_i^\source, y_i^\source)\}_{i=1}^n$, are concatenated into $(\bX^\source, \by^\source)\in \R^{n \times d} \times \R^n$.
Among the $m$ targets data, $m_k$ are labeled, $\{(\bx_i^\target, y_i^\target)\}_{i=1}^{m_k}$, and concatenated into $(\bX^{\target_\ell}, \by^{\target_\ell}) \in \R^{m_k \times d} \times \R^{m_k}$.
Overall, $(\bX^\source, \by^\source, \bX^{\target_\ell}, \by^{\target_\ell})$ constitutes the training set.
The remaining target data $\{\bx_i^\target\}_{i = m_k+1}^{m}$ are unlabeled, and the goal is to predict their associated outcomes $\{\by_i^\target\}_{i = m_k+1}^{m}$.

\subsection{Learning problem}

\textbf{Overall loss}
Based on models from section~\ref{sec:problem}, we propose a loss function to jointly estimate $\bU^\source, \bU^\target$, and $\bbeta$ for regression and classification problems.
Given $\cOT, \cgrass \geq 0$, the overall loss is
\begin{equation}
    \label{eq:loss}
    \loss(\bU^\source, \bU^\target, \bbeta, \bpi) \triangleq \losssup(\bU^\source, \bU^\target, \bbeta) +  \cOT \lossemb(\bU^\source, \bU^\target, \bpi)  +  \cgrass \lossgrass(\bU^\source, \bU^\target).
\end{equation}
This loss is a sum of three terms: an empirical risk $\losssup$ on the score functions,
a metric learning loss $\lossemb$ between the embeddings $(\bU^\source)^\top \bx_i^\source$ and $(\bU^\target)^\top \bx_i^\target$ and a regularization loss $\lossgrass$ to control how far $\spann(\bU^\target)$ is from $\spann(\bU^\source)$.

\textbf{Empirical risk $\losssup$}
The loss function on the score function is a regularized empirical risk on the source data and the few labeled target data
\begin{equation}
    \label{eq:loss_dec}
    \losssup(\bU^\source, \bU^\target , \bbeta) \triangleq \sum_{\domain \in \{\source, \target_\ell \}} \sum_{i=1}^{|\domain|} \mathcal{L}\left(y_i^\domain, \bbeta^\top (\bU^\domain)^\top \bx_i^\domain + \beta_0 \right)+ \csup \norm{\bbeta}_2^2
\end{equation}
where $\mathcal{L}: \R \times \R \to \R$, and $\csup \geq 0$.
For a regression problem, we leverage the classic square error $\mathcal{L}(y, \hat{y}) = (y - \hat{y})^2$.
For a binary classification problem in $\{-1, 1\}$, the chosen    loss function is the logistic loss $\mathcal{L}(y, \hat{y}) = \log\left(1 + \exp( - y \hat{y} ) \right)$.

\textbf{Metric learning loss $\lossemb$}
Then, we introduce the metric learning loss that enhances the estimation of $\bU^\source$ and $\bU^\target$ leveraging a geometrical point of view.
First, we leverage optimal transport (OT)~\cite{peyre_computational_2019, grave_unsupervised_2019} to estimate a mapping $\bpi \in \bPi(n, m)$ between the source and target domains and hence bring closer points from the domains with a high value of $\pi_{ij}$.
Second, we compute intra-domain affinity matrices $\bK(\by)$ so that two points with similar $y_i$ should be brought closer.
Overall, this metric learning loss function is 
\begin{equation}
    \begin{multlined}
        \lossemb(\bU^\source, \bU^\target, \bpi) \triangleq \overbrace{\langle \bC(\bX^\source\bU^\source, \bX^\target\bU^\target), \bpi \rangle}^{\textup{OT loss}} \\
        + \underbrace{\sum_{\domain \in \{\source, \target_\ell\}}  \frac{|\domain|}{n+m_k} \langle \bC(\bX^\domain\bU^\domain, \bX^\domain\bU^\domain), \bK(\by^\domain) \rangle}_{\textup{similarity loss}}
    \end{multlined}
\end{equation}
where $\langle \bXi, \bEta \rangle \triangleq \Tr(\bXi^{\top}\bEta)$,  $(\bC(\bX, \bX^\prime))_{ij} \triangleq \norm{\bx_i - \bx_j'}_2^2$ and $\bK(\by)$ is an affinity matrix: for a regression problem, we leverage the normalized Gaussian kernel whose elements are $(\bK(\by))_{ij} =  \exp(-\frac{1}{2}(y_i - y_j)^2)/\sum_{l,m} \exp(-\frac{1}{2}(y_l - y_m)^2)$; for classification problems, we use the matrix with elements  $(\bK(\by))_{ij} =  1_{y_i = y_j}/\sum_{l,m} 1_{y_l = y_m}$.
It should be noted that the joint distribution $\bpi$ estimates the permutation matrix from the assumption of equation~\eqref{eq:pi}.

\textbf{Grassmann loss $\lossgrass$}
We now move on to the last term of the total loss~\eqref{eq:loss}.
Unfortunately, learning $\bU^\target$ from data seems infeasible since we have few labels in the target domain.
To alleviate this problem, we regularize $\spann(\bU^\source)$ and $\spann(\bU^\target)$ to be not too far apart.
To do so, we leverage the Riemannian distance of the Grassmann manifold~\cite{edelman_geometry_1998}, i.e. 
\begin{equation}
    \begin{aligned}
        \lossgrass(\bU^\source, \bU^\target)  &\triangleq \frac{1}{\sqrt{2}} \norm{\bU^\source(\bU^\source)^{\top} - \bU^\target(\bU^\target)^{\top}}_F\\
        &= \left[\sum_{i=1}^q \sin(\theta_i)^2\right]^{\nicefrac{1}{2}}
    \end{aligned}
\end{equation}
where $\theta_1, \dots, \theta_k$ are the principle angles between $\bU^\source$ and $\bU^\target$~\cite{edelman_geometry_1998}.
It's important to note that as $\cgrass \to +\infty$ in the overall loss~\eqref{eq:loss}, $\spann(\bU^\target) \to \spann(\bU^\source)$, and hence $\bU^\target \to \bU^\source \bO$ for some $q\times q$ orthogonal matrix $\bO$.
Thus, for $q$ much lower than $d$, $\lossgrass$ controls the complexity to estimate $\bU^\target$ as desired.

\subsection[Reformulation as a problem on the Stiefel manifold and optimization]{Reduction to a problem on the Stiefel manifold $\St^2$ and optimization}

\begin{algorithm}[tb]
    \caption{Alternate optimization}
    \label{alg:optim}
    \begin{algorithmic}[1]
        \STATE Initialize: $(\bU^\source, \bU^\target) \in \St^2$
        \REPEAT
            \STATE Compute $\bpi^*$ and $\bbeta^*$ as in~\eqref{eq:min_pi_beta}
            \STATE Perform a Riemannian gradient descent step on $\St^2$ of $\lossSt$~\eqref{eq:min_problem}
        \UNTIL{convergence}
    \end{algorithmic}
\end{algorithm}

We now discuss the optimization of the loss function~\eqref{eq:loss}.
The optimization problem is written
\begin{equation}
    \label{eq:min_problem_1}
    \minimize_{(\bU^\source, \bU^\target, \bbeta, \bpi) \in \St^2 \times \R^q \times \bPi(n, m)} \loss(\bU^\source, \bU^\target, \bbeta, \bpi).
\end{equation}
Given $(\bU^\source, \bU^\target) \in \St^2$ fixed, $\loss$ is easily minimized with respect to $(\bbeta, \bpi) \in \R^q \times \bPi(n, m)$.
Indeed, denoting
\begin{equation}
    \label{eq:min_pi_beta}
    \left(\bbeta^*, \bpi^*\right) = \argmin_{(\bbeta, \bpi) \in \R^q \times \bPi(n, m)} \loss(\bU^\source, \bU^\target, \bbeta, \bpi),
\end{equation}
$\bbeta^*$ is computed with a ridge or logistic solver and $\bpi^*$ with an OT solver.
Hence,~\eqref{eq:min_problem_1} reduces to
\begin{equation}
    \label{eq:min_problem}
     \minimize_{(\bU^\source, \bU^\target) \in \St^2}  \bigg\{  \lossSt(\bU^\source , \bU^\target )  \triangleq  \loss(\bU^\source , \bU^\target , \bbeta^* , \bpi^* )  \bigg\}.
\end{equation}
The latter is solved with a Riemannian gradient descent algorithm~\cite{absil_optimization_2008, boumal_introduction_2023} leveraging a Riemannian Adam optimizer~\cite{becigneul_riemannian_2019} on the Stiefel manifold~\cite{manton_optimization_2002}.
This alternate optimization procedure is summed up in algorithm~\ref{alg:optim}. 

\subsection{Related work}

\textbf{Unsupervised subspace learning}
Numerous methods to address domain shifts exist~\cite{zhuang_comprehensive_2021}.
Focusing on subspace-based methods, Transfer Component Analysis (TCA)~\cite{pan_domain_2011} learns some transfer components across domains in a reproducing kernel Hilbert space using maximum mean discrepancy to align source and target distributions.
Other methods such as Subspace Alignment~\cite{fernando_unsupervised_2013}, joint cross-domain classification and subspace learning for unsupervised adaptation~\cite{fernando_joint_2015} and principle components-based methods~\cite{maman_domain_2019} learn source components leveraging target principal components to align data from both domains.
These methods are unsupervised in the sense they do not use the target labels.
In contrast, MSA benefits from target labels to simultaneously learn embeddings, with the two Stiefel matrices, and the predictor, which helps with low signal-to-noise ratio M/EEG applications.
Indeed,~\cite{mellot_harmonizing_2023} showed the inefficiency of unsupervised domain adaptation methods such as~\cite{maman_domain_2019} for brain age applications.

\textbf{Semi-supervised rotation correction}
Methods tailored for domain adaptation in M/EEG applications have emerged alongside domain-specific whitening~\eqref{eq:embedding}.
Riemannian Procrustres Analysis (RPA)~\cite{rodrigues_riemannian_2019} aligns covariance matrix class centers by learning with a gradient descent a $p \times p$ orthogonal matrix, involving $\mathcal{O}(p^2)$ parameters.
In the same spirit, Tangent Space Alignment (TSA)~\cite{bleuze_transfer_2021} aligns tangent space vector class centers with a $d \times d$ orthogonal matrix, requiring $\mathcal{O}(p^4)$ parameters, learned through singular value decomposition.
These methods apply only to classification datasets.
In comparison, MSA learns two Stiefel matrices, that is, $\mathcal{O}(p^2 q)$ parameters, where $q$ is user-defined.
This hyperparameter has the benefit of controlling the cost computation and the complexity of the model depending on the task.
Furthermore, it applies to both regression and classification tasks.

\section{Numerical experiments}
    \label{sec:num_exp}
    In this section, we assess the performance of MSA on a regression problem on MEG data.
This problem is subject-specific, i.e., each subject has one outcome.
We have two sets of subjects without any overlap from datasets where a shift has occurred, such as variations in tasks performed by the subjects.
In order to promote research reproducibility, code is available on github\footnote{\href{github.com/antoinecollas/MSA}{https://github.com/antoinecollas/MSA}}.

\subsection{Regression on MEG data}
\label{subsec:meg}

\textbf{Cam-CAN dataset}
We utilize the well-established Cambridge Center of Aging and Neuroscience (Cam-CAN) dataset~\cite{taylor_cambridge_2017}.
Our objective is to predict the age of individuals based on their MEG
recordings, a regression problem of significant interest in neuroscience research.
Indeed, results from~\cite{kaufmann_common_2019} suggest that the prediction error of models trained to learn age from brain data of healthy populations provides clinically relevant information related to neurodegenerative anomalies physical and cognitive decline.
In this dataset, $646$ subjects ($319$ female, $327$ male) were collected with the same $306$-channel VectorView MEG system (Elekta Neuromag, Helsinki) at $1$kHz frequency.
Their age distribution is from $18.5$ to $88.9$ years with an average of $54.9 \pm 18.4$ years with an almost uniform spread over the age range.
The dataset is divided into two sets of subjects with no overlap but with the same age distribution.
Notably, the subjects perform different tasks, depending on the domain, during data acquisition.
These tasks are rest (eyes closed), passive (audio-visual stimuli), and somatosensory tasks (audio-visual plus a manual response).
Hence, we get one time series and one outcome, the age, per subject.
Preprocessing is applied to these time series; see Appendix~\ref{sec:preprocessing} for the details.
To get the semi-supervised setting, $10\%$ of the target labels are kept in addition to the source labels during training.
Then, we predict ages with several methods, all involving ridge regression.

\begin{figure*}
    \centering
    \includegraphics[width=\linewidth]{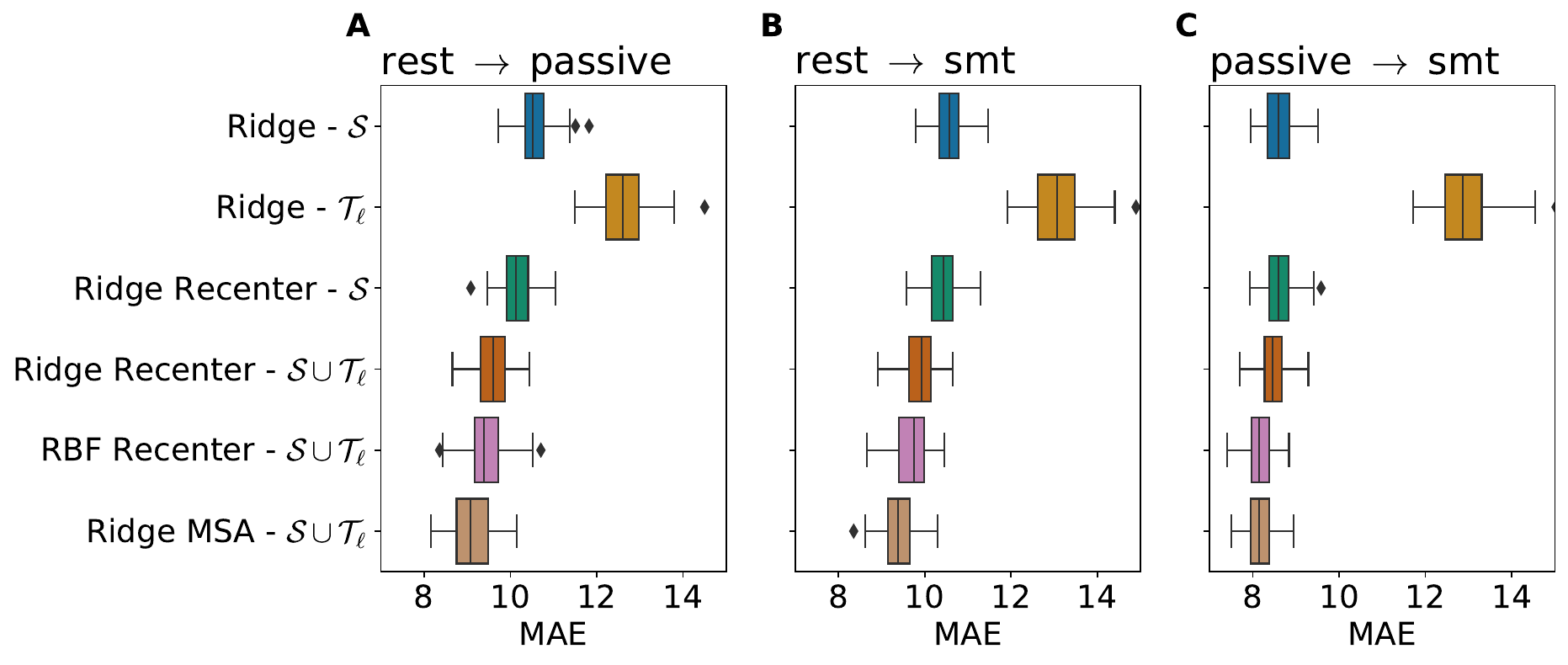}
    \caption{
        \textbf{Mean absolute errors (MAE) of different regressors on the brain age prediction problem of the Cam-CAN dataset (the lower, the better).}
        The $646$ subjects ($p=65$) are split into source and target domains ($323$ subjects each) and are associated with two different tasks.
        The latter are reported over each subfigure, e.g., the subfigure (\textbf{A}) presents results with the rest for the source task and the passive task for the target.
        $10\%$ of the target labels are kept during training, and $100$ different data splits are performed.
    }
    \label{fig:meg_perfs}
\end{figure*}

\begin{figure}
    \centering
    \includegraphics[width=\linewidth]{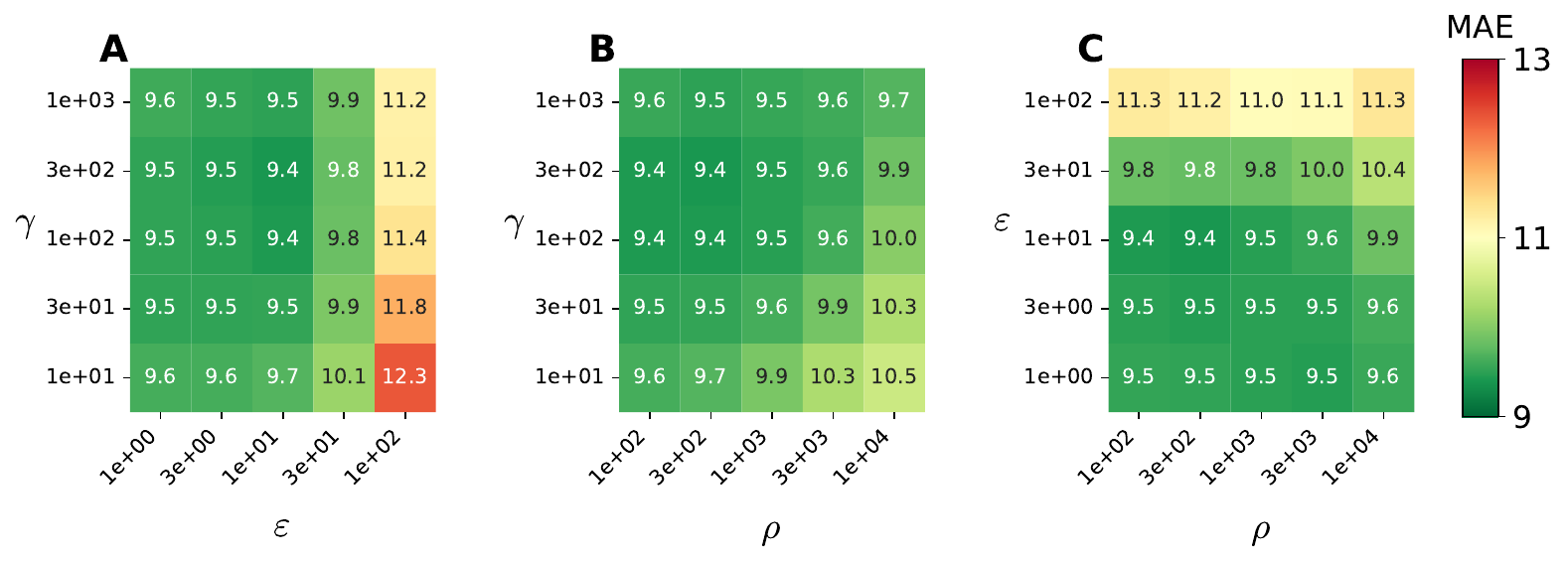}
    \caption{
        \textbf{Mean absolute errors (MAE) of MSA for different values of hyperparameters on the brain age prediction problem of the Cam-CAN dataset (the lower, the better).}
        The source task is rest, and the target task is somatosensory.
        $10\%$ of the target labels are kept during training, and $100$ different data splits are performed.
    }
    \label{fig:meg_hyp_sensitivity}
\end{figure}

\begin{figure}
    \centering
    \includegraphics[width=0.7\linewidth]{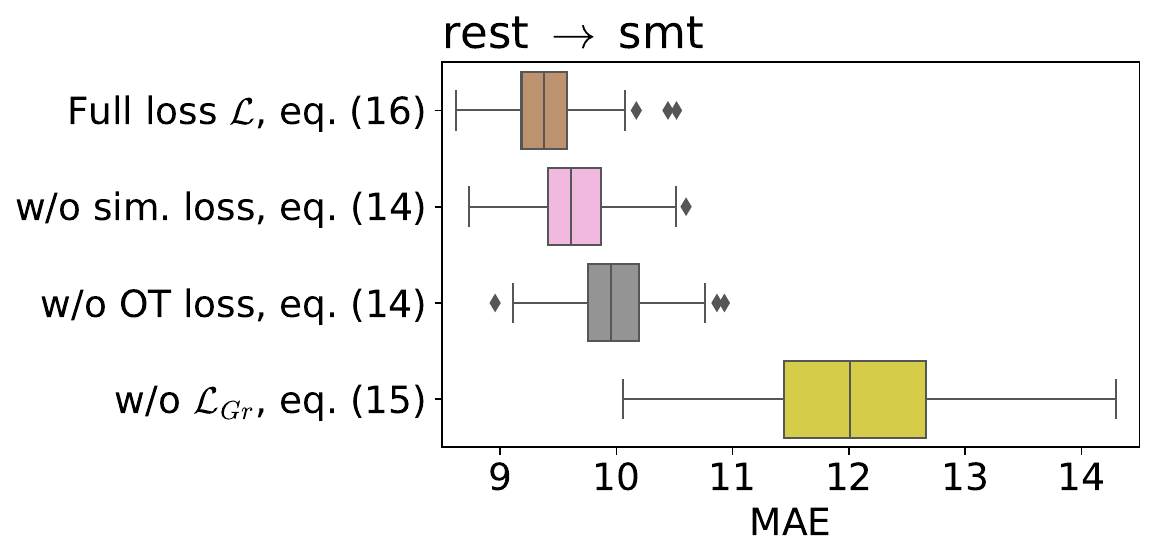}
    \caption{
        \textbf{Ablation study of the proposed loss~\eqref{eq:loss} on the brain age prediction problem of the Cam-CAN dataset.}
        Mean absolute errors (MAE) of the full loss versus when one of the terms is removed are reported (the lower, the better).
        The source task is rest, and the target is somatosensory.
        $10\%$ of the target labels are kept during training, and $100$ different data splits are performed.
    }
    \label{fig:meg_ablation}
\end{figure}

\textbf{Methods}
First, we compare MSA with ridge regressions that are trained on embeddings either of the source domain or of the $10\%$ labeled target domain, as in~\cite{sabbagh_manifold-regression_2019}.
These embeddings~\eqref{eq:embedding} are computed at the Riemannian mean~\eqref{eq:mean} of the full dataset.
We denote these two methods \texttt{Ridge - $\source$} and \texttt{Ridge - $\target_\ell$} respectively.
Then, we leverage recently proposed methods for brain age prediction under domain shifts: ridge regressions are trained either on the source domain or on the concatenation of the source and labeled target domains with embeddings with domain-specific Riemannian means as proposed in~\cite{mellot_harmonizing_2023}.
They are denoted \texttt{Ridge Recenter - $\source$} and \texttt{Ridge Recenter - $\source \cup \target$} respectively.
Recently,~\cite{bonet_sliced-wasserstein_2023} proposed to use kernel ridge regression with an RBF kernel for brain age prediction and showed state-of-the-art results.
One kernel per frequency band is computed with embeddings, i.e., $K_{i,j} = \exp(-\frac{1}{2\sigma^2} \norm{\bx_i - \bx_j}_2^2)$ for $\bx_i,\bx_j$ in $\bX^\source$ and/or $\bX^{\target_\ell}$ and where $\sigma^2$ is the hyperparameter of length scale.
Then, kernels are summed to get a single kernel.
All labeled data (source and labeled target) are used to compute the kernels.
We denote this method \texttt{RBF Recenter - $\source \cup \target_\ell$}.
Finally, MSA is denoted \texttt{Ridge MSA - $\source \cup \target_\ell$} and is used with rank $q=30$.
Performance with the best-performing hyperparameters is reported for all methods.

\textbf{Performance comparison}
Box plots of mean absolute errors (MAE) on the $90\%$ unlabeled data are reported in Figure~\ref{fig:meg_perfs} with training and testing on $100$ different data splits.
First of all, we observe that training solely on the target set (\texttt{Ridge - $\target_\ell$}) leads to the worst MAE.
Hence, applying the method of~\cite{sabbagh_manifold-regression_2019} is not recommended when too few labels are available.
Then, \texttt{Ridge Recenter - $\source$} gives lower MAE than \texttt{Ridge - $\source$}, which shows the importance of considering domain adaptation techniques when considering changes of tasks.
The three other methods, \texttt{Ridge Recenter - $\source \cup \target$}, \texttt{RBF Recenter - $\source \cup \target_\ell$} and \texttt{Ridge MSA - $\source \cup \target_\ell$} all leverage these domain adaptation techniques and the few labels from the test set.
Since they use these additional labels, they outperform the other methods.
The \texttt{RBF Recenter - $\source \cup \target_\ell$} is the only non-linear model and performs slightly better than \texttt{Ridge Recenter - $\source \cup \target$}.
Overall, \texttt{Ridge MSA - $\source \cup \target$} performs better than all other methods, regardless of the considered pair of tasks.

\textbf{Parameters sensitivity and ablation study}
Then, we study the hyperparameter sensitivity of MSA.
The three considered hyperparameters are $\gamma$, $\rho$ and $\varepsilon$ from the loss~\eqref{eq:loss}.
$100$ data splits are performed for each triplet value, and MAE is reported in Figure~\ref{fig:meg_hyp_sensitivity}.
This shows a certain robustness to the choice of hyperparameters.
Finally, to assess the need for the different losses in the overall loss~\eqref{eq:loss}, we ablate each one, search for the best-remaining hyperparameters $\gamma$, $\rho$, and $\varepsilon$, and report MAE in Figure~\ref{fig:meg_ablation}.
We observe that each ablation leads to poorer MAE, which empirically justifies using all terms together in the full loss~\eqref{eq:loss}.
In particular, removing the Grassmann loss drops performance a lot because it resumes to estimate two different Stiefel matrices.
Notably, removing the optimal transport also leads to a high-performance drop, which motivates the ``OTDA assumption" introduced in section~\ref{subsec:mixing}.

\section{Conclusions}
In conclusion, this paper has introduced a new domain adaptation approach for time series data under mixing models.
The key innovation lies in utilizing a domain-dependent mixing matrix~\eqref{eq:mixing_model_compact} and the OTDA assumption~\eqref{eq:pi}.
Indeed, we identified two Stiefel matrices that, applied to a Riemannian representation of observed signal covariances, recover variances from the underlying signal.
An integrated cost function enabled simultaneous learning of these matrices, establishing pairwise relationships between source and target domains and constructing task-specific predictors.
Applied to MEG problems, this approach has outperformed recent methods in brain-age regression.

\section{Acknowledgments}
The scientific Python ecosystem enabled numerical computation:
Geoopt~\cite{kochurov_geoopt_2020},
Matplotlib~\cite{hunter_matplotlib_2007},
Numpy~\cite{harris_array_2020},
POT~\cite{flamary_pot_2021},
Pyriemann~\cite{barachant_pyriemannpyriemann_2023},
Pytorch~\cite{paszke_pytorch_2019},
Seaborn~\cite{waskom_seaborn_2021},
and Sklearn~\cite{pedregosa_scikit-learn_2011}.

\newpage
\bibliographystyle{plain}
\bibliography{references}

\newpage
\appendix
\section{Technical details}
\label{sec:appendices}

\subsection{Proof of Proposition~\ref{prop:projections}}
\label{app:proof}

Without loss of generality, we assume that $\bpi$ is the identity, and thus, from equation~\eqref{eq:pi} $\diag(\bp_i^\source) = \diag(\bp_i^\target)$.
Given $\bW^\source = (\bSigmaMean^\source)^{\nicefrac{-1}{2}} \bA^\source (\bEMean^\source)^{\nicefrac{1}{2}} \in \Ort(p)$, we have
\begin{equation*}
    \log\left( (\bSigmaMean^\source)^{\nicefrac{-1}{2}} \bSigma_i^\source (\bSigmaMean^\source)^{\nicefrac{-1}{2}} \right) = \bW^\source \log\left( (\bEMean^\source)^{\nicefrac{-1}{2}} \bE_i^\source (\bEMean^\source)^{\nicefrac{-1}{2}} \right) (\bW^\source)^{\top}.
\end{equation*}
This implies that
\begin{align*}
    \bx_i^\source &= \vect\left(\log\left( (\bSigmaMean^\source)^{\nicefrac{-1}{2}} \bSigma_i^\source (\bSigmaMean^\source)^{\nicefrac{-1}{2}} \right)\right) \\
    &= (\bW^\source \otimes \bW^\source) \vect \left(\log\left( (\bEMean^\source)^{\nicefrac{-1}{2}} \bE_i^\source (\bEMean^\source)^{\nicefrac{-1}{2}} \right)\right).
\end{align*}
Since $\log\left( (\bEMean^\source)^{\nicefrac{-1}{2}} \bE_i^\source (\bEMean^\source)^{\nicefrac{-1}{2}} \right)$ is a block diagonal matrix whose upper $q \times q$ block is $\diag\left(\log(p_{i, 1}/\pbar_1), \dots, \log(p_{i, q}/\pbar_q)\right)$, there exists a basis $\bQ = \left[\bU, \bU_\perp\right] \in \Ort(d)$ such that
\begin{equation*}
    \vect \left(\log\left( (\bEMean^\source)^{\nicefrac{-1}{2}} \bE_i^\source (\bEMean^\source)^{\nicefrac{-1}{2}} \right)\right) = \bU \left[\log(p_{i, 1}/\pbar_1), \dots, \log(p_{i, q}/\pbar_q) \right]^{\top} + \bU_\perp \bn_i^\source
\end{equation*}
where $\bn_i^\source = \bU_\perp^{\top}\vect \left(\log\left( (\bEMean^\source)^{\nicefrac{-1}{2}} \bE_i^\source (\bEMean^\source)^{\nicefrac{-1}{2}} \right)\right)$.
It follows that for $\bU^\source \triangleq (\bW^\source \otimes \bW^\source) \bU$, we have
\begin{equation*}
    (\bU^\source)^{\top} \bx_i^\source = \left[\log(p_{i, 1}/\pbar_1), \dots, \log(p_{i, q}/\pbar_q) \right]^{\top}.
\end{equation*}
Applying the same reasoning to the target embeddings, we get that for $\bU^\target \triangleq (\bW^\target \otimes \bW^\target) \bU$, we have $(\bU^\target)^{\top} \bx_i^\target = \left[\log(p_{i, 1}/\pbar_1), \dots, \log(p_{i, q}/\pbar_q) \right]^{\top}$.

\section{Numerical experiments details}

\subsection{Datasets preprocessing}
\label{sec:preprocessing}

We follow the preprocessing steps proposed in~\cite{engemann_reusable_2022, mellot_harmonizing_2023} for the Cam-CAN dataset.
We apply a FIR band-pass filter between $0.1$ and $49$Hz.
We decimate the signals to get a sampling frequency of $200$Hz.
To compensate for environmental noise, we perform a temporal signal-space-separation (tSSS) method~\cite{taulu_applications_2005} with a chunk duration of $10$ seconds and a correlation threshold of $98$\%.
Then, we retain channels corresponding to magnetometers.
Each filtered recording is segmented in $10$s epochs without overlap.
Then, epochs are filtered into $7$ frequency bands: $0.1$-$1$, $1$-$4$, $4$-$8$, $8$-$15$, $15$-$26$, $26$-$35$, and $35$-$49$Hz.
We performed artifact rejection by thresholding extreme peak-to-peak amplitudes on single epochs using the local autoreject method~\cite{jas_autoreject_2017}.
We average the epochs and compute one covariance matrix per subject and per frequency band using the Oracle Approximating Shrinkage (OAS) estimator~\cite{chen_shrinkage_2010}.
Then, principal component analysis transforms them to matrices of size $p=65$ .
We compute embeddings $\bx_i^\domain$, still for each frequency band, using the function $\phi$ from equation~\eqref{eq:vectorize}.
Finally, these embeddings of all frequency bands are concatenated into one embedding per subject.

\subsection{MEG additional figures}

\begin{figure*}[h]
    \centering
    \includegraphics[width=\linewidth]{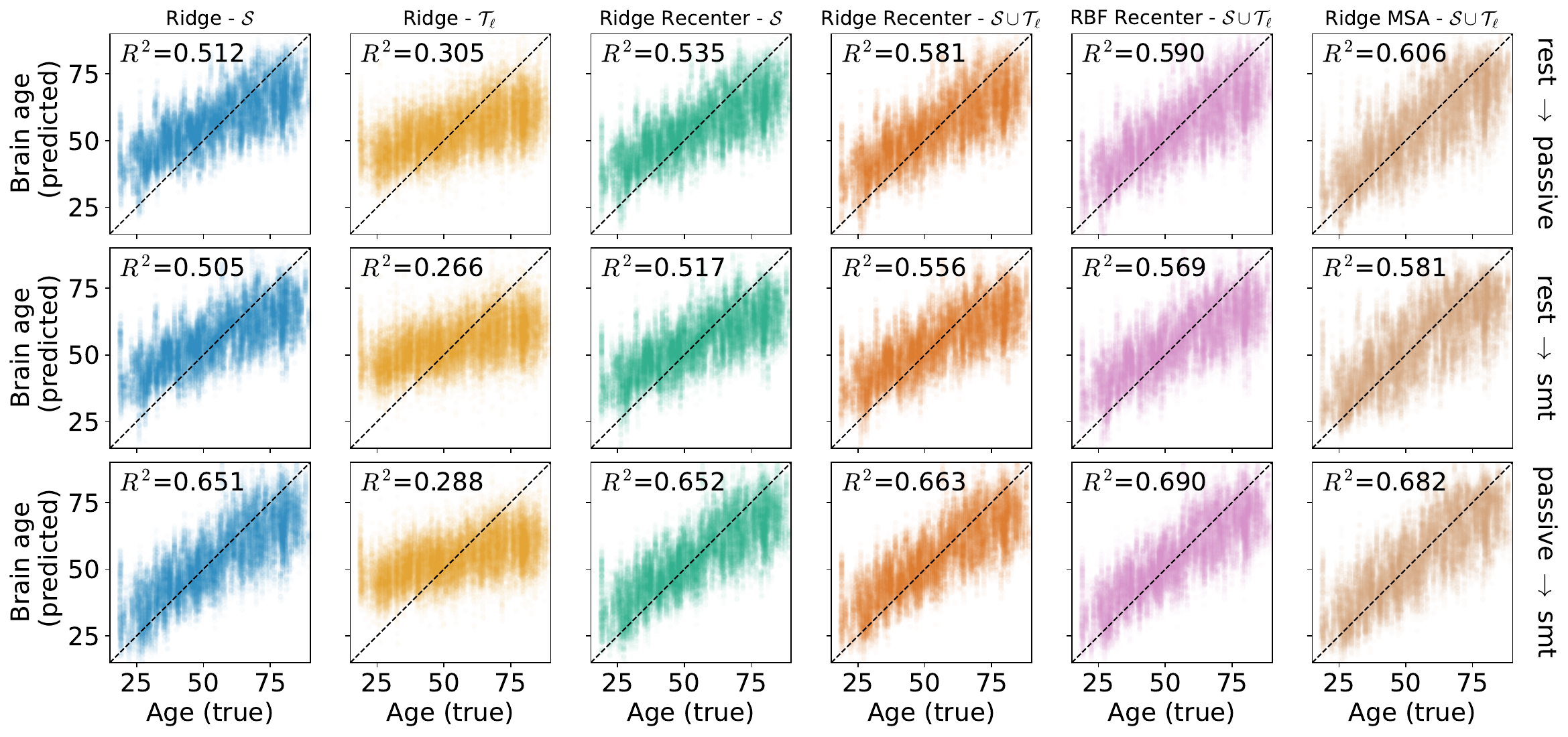}
    \caption{
        \textbf{Scatter plots of different regressors on the brain age prediction problem of the Cam-CAN dataset.}
        $R^2$ scores are reported for each method on each pair of tasks (the higher, the better).
    }
\end{figure*}

\begin{figure*}
    \centering
    \includegraphics[width=\linewidth]{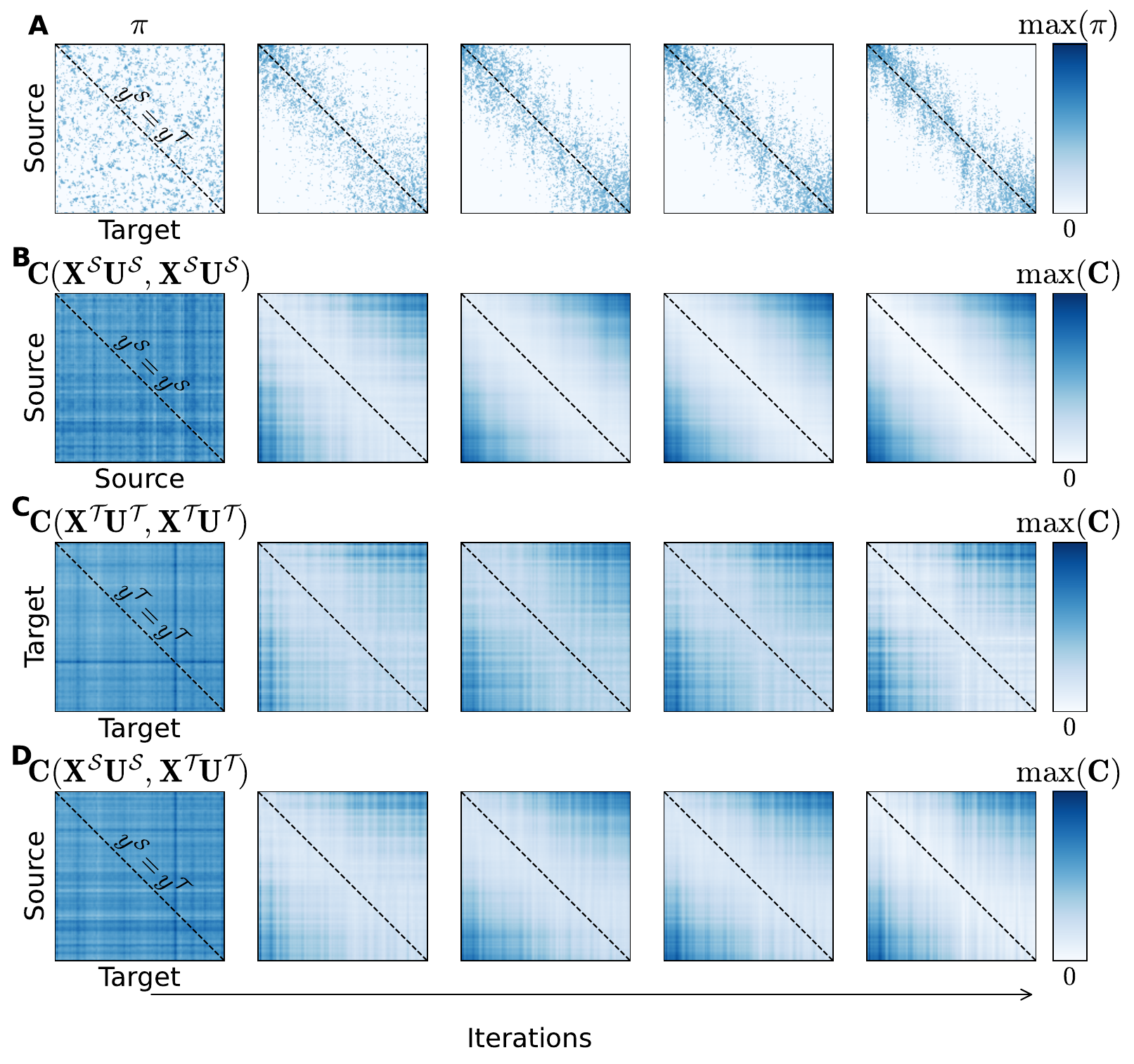}
    \caption{
        \textbf{Evolutions of different quantities involved in the loss~\eqref{eq:loss} on the brain age prediction problem of the Cam-CAN dataset.}
        The source task is rest, and the target task is somatosensory.
        Each quantity is computed $100$ times with different data splits and then averaged.
    }
\end{figure*}

\begin{figure*}
    \centering
    \includegraphics[width=0.5\linewidth]{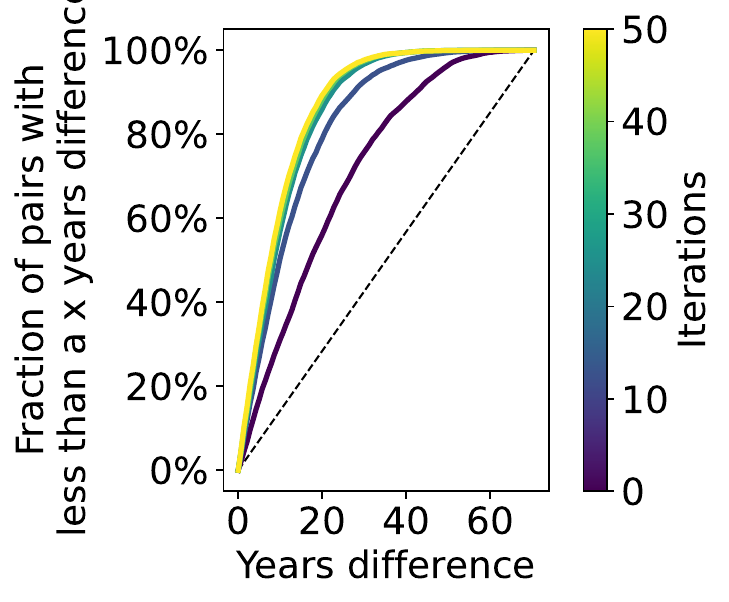}
    \caption{
        \textbf{Fraction of pairs $\pi_{ij}$ with less than a given number of years difference on the brain age prediction problem of the Cam-CAN dataset for different numbers of iterations of the Algorithm~\ref{alg:optim}.}
        The source task is rest, and the target task is somatosensory.
        Given a number of iterations, the optimal transport plan $\bpi$ is computed $100$ times with different data splits and then averaged.
    }
\end{figure*}

\end{document}